\documentclass[sigconf,authorversion,nonacm]{acmart}

\setcopyright{none}
\renewcommand\footnotetextcopyrightpermission[1]{} % removes footnote with conference information in first column

\usepackage{xcolor}

\usepackage{fix-cm}
\usepackage{textcomp}
\usepackage{tlatex}
\usepackage[frozencache,cachedir=minted-cache]{minted}
\usepackage{svg}
\svgsetup{inkscapelatex=false}
\usepackage{listings}
\newcommand{\Etime}{\emph{ET}}
\newtheorem{thm}{Theorem}

\AtBeginDocument{%
  }

\setcopyright{acmlicensed}
\acmYear{2026}
\begin{document}

\title{LeaseGuard: Raft Leases Done Right}
\titlenote{This paper has been accepted to ACM SIGMOD 2026. This is the author-accepted manuscript.}

\author{A. Jesse Jiryu Davis}
\orcid{0009-0002-0909-2207}
\email{jesse@mongodb.com}
\affiliation{%
  \institution{MongoDB Research}
    \city{New York City}
  \state{NY}
  \country{USA}
}
\author{Murat Demirbas}
\orcid{0000-0001-7952-9035}
\email{murat.demirbas@mongodb.com}
\affiliation{%
  \institution{MongoDB Research}
    \city{New York City}
  \state{NY}
  \country{USA}
}
\author{Lingzhi Deng}
\orcid{0009-0002-2072-3989}
\email{lingzhi.deng@mongodb.com}
\affiliation{%
  \institution{MongoDB, Inc.}
    \city{New York City}
  \state{NY}
  \country{USA}
}

\renewcommand{\shortauthors}{Davis et al.}
\begin{abstract}
Raft is a leading consensus algorithm for replicating writes in distributed databases. However, distributed databases also require consistent \textit{reads}. To guarantee read consistency, a Raft-based system must either accept the high communication overhead of a safety check for each read, or implement \textit{leader leases}. Prior lease protocols are vaguely specified and hurt availability, so most Raft systems implement them incorrectly or not at all.
We introduce LeaseGuard, a novel lease algorithm that relies on guarantees specific to Raft elections. LeaseGuard is simple, rigorously specified in TLA+, and includes two novel optimizations that maximize availability during leader failover. The first optimization restores write throughput quickly, and the second improves read availability.
We evaluate LeaseGuard with a simulation in Python and an implementation in LogCabin, the C++ reference implementation of Raft. By replacing LogCabin's default consistency mechanism (quorum checks), LeaseGuard reduces the overhead of consistent reads from one to zero network roundtrips. It also improves write throughput from $\sim$1000 to $\sim$10,000 writes per second, by eliminating contention between writes and quorum reads. Whereas traditional leases ban all reads on a new leader while it waits for a lease, in our LeaseGuard test the new leader instantly allows 99\% of reads to succeed.
\end{abstract}



% \keywords{Lease, Raft, Clock, Consensus}

\received{17 December 2025}

\maketitle

% Tell LaTeX how to split certain words
\hyphenation{Commit-Entry}
\hyphenation{Client-Read}
\hyphenation{commitIndex}

\section{Introduction}

The Raft consensus algorithm is used in MongoDB~\cite{raftMDB}, CockroachDB~\cite{crdb}, TiDB~\cite{tidb}, YugabyteDB~\cite{yugabyte}, and others~\cite{etcd,rethink,consul}. It reliably replicates a log of write commands and handles leader failures. But replication introduces the risk of reading from a stale replica. If a database served stale reads while promising strong consistency, it would break its promise. For example, CockroachDB and YugabyteDB promise serializability and strongly consistent reads by default; TiDB and MongoDB support snapshot isolation. Violating these guarantees would lead to application misbehavior~\cite{googleStrongConsistency}, lost updates~\cite{JimGrayLectureNotes}, and security vulnerabilities~\cite{acidrain}, so the database needs a mechanism to ensure it reads from a fresh replica.

It is surprisingly hard to ensure consistent reads from a Raft leader. The leader is elected at the start of a \textit{term} and executes all reads and writes until it crashes or is \textit{deposed}, i.e., a new leader is elected with a higher term number. The deposed leader may be unaware for some time that it has been deposed, for example if it is on the minority side of a network partition. If it continues serving reads while the new leader executes writes, the deposed leader returns stale data, violating consistency. Although elections may be rare in a single Raft group, large deployments make ``rare'' events common. MongoDB's public DBaaS fleet has about 200,000 Raft replica sets, of which half experience an election each week. Half of those elections are related to routine maintenance, and half are unplanned. A system of this size requires some mechanism to avoid stale reads during elections. Unfortunately, Raft's original mechanism~\cite{raft} is expensive: before responding to each read request, a leader exchanges messages with a majority of the replica set to confirm it is still the highest-term leader. This quorum check incurs latency, I/O contention, and monetary costs in the cloud. It is a high price to pay for every query, just to ensure consistency during occasional elections.

Leader leases have been proposed~\cite{ongaroThesis} to make Raft reads consistent and efficient. A lease ensures that at most one node believes it is the leader at a time, so it can serve consistent reads locally without talking to other nodes. Unfortunately, the original description of Raft leases is vague, without code or formal specifications (see Section~\ref{sec:relwork}). The same applies to leader leases in Paxos, Raft's predecessor~\cite{paxosLive,spanner}.
For this reason, Raft systems often implement leases with bugs. HashiCorp's Raft implementation, for example, does not follow the Raft paper. It can violate consistency due to message delays, or because a new leader does not know of a prior leader's lease~\cite{consulLease,jepsenEtcdConsul}. These bugs were reported in 2016. The etcd Raft implementation also allows stale reads due to miscalculations of lease timeouts~\cite{boringhello}. This bug was reported in 2024. All these bugs were still open at the time of writing. Most Raft implementations forego leases. They use expensive quorum checks, or they do not guarantee consistency at all. MongoDB's Raft variant~\cite{raftMDB} avoids leases: if a user demands consistent reads, the leader writes an empty entry to its log for \textit{each} read request, and waits for majority replication of the entry to confirm it is not deposed. 

Prior lease designs also hurt availability: a deposed leader's lease must expire before a new leader can process reads and writes. (E.g., TiDB's Raft lease causes 10 seconds of unavailability after an election~\cite{timestampAsAService}.) During this period, blocked operations may queue up, then overwhelm the leader as soon as it acquires a lease; and this thundering herd could in turn cause a metastable failure~\cite{metastable}. Prior designs also risk gray failures, where a leader may continue renewing its lease, but fail to execute writes due to internal issues such as disk failure. The system is stuck with a ``faux leader'' that cannot make progress yet refuses to step aside. 

LeaseGuard is a new lease protocol for Raft which solves the problems above. It is specified in detail, and it is simpler than previous lease designs: there are no changes to Raft messages, no additional messages, no new data structures, and no changes to the election protocol. Instead, in LeaseGuard \textbf{the log is the lease}. The leader establishes its lease by committing a log entry, and followers learn of the lease by replicating the log normally. We can rely on the log due to a Raft-specific guarantee called \textbf{Leader Completeness}: a newly elected leader is guaranteed to have all log entries that were replicated by a majority in the previous term. (Competing consensus protocols like Paxos and Viewstamped Replication~\cite{Liskov2012} do not guarantee Leader Completeness.) This guarantee enables a simpler lease design, reduces implementation bugs, and enables availability optimizations overlooked in prior work. It also solves the gray-failure/faux-leader problem: only a leader who can make real progress (by writing and replicating log entries) can maintain its lease.

It is critical for distributed databases to maintain high availability and reliable performance, and avoid metastable failures.
LeaseGuard minimizes write unavailability through its {\bf deferred-commit writes} optimization (Section~\ref{sec:deferred-commit-writes}), allowing the new leader to write and replicate log entries while it waits for the deposed leader's lease to expire.
LeaseGuard also minimizes read unavailability through its {\bf inherited lease reads} optimization, enabling a new leader to serve some consistent reads concurrently with a deposed leader during the deposed leader's lease period, with zero communication overhead. This optimization in particular requires bounded-uncertainty clocks. Such clocks have recently become available in the cloud.
Our formal TLA+~\cite{tlaplus} specification and correctness proofs (Section~\ref{sec:correct}) validate the safety of these optimizations. Notably, we discovered these surprising optimizations through our exploration of leases in TLA+.
Our improvements to read availability and write throughput together decrease the risk that a thundering herd of operations will overwhelm a new leader as soon as it acquires a lease.

To evaluate performance, we developed a Python simulation\footnote{https://github.com/muratdem/RaftLeaderLeases/tree/main/Python} of Raft and LeaseGuard (Section~\ref{sec:simulation}). The simulator models nodes as Python objects that communicate asynchronously, similar to Python's \texttt{asyncio} library. It enables deterministic, reproducible experiments given a set of parameters and a random seed. We also implemented a linearizability checker to validate protocol behavior.

For real-world validation, we implemented LeaseGuard in C++\footnote{
https://github.com/mongodb-labs/logcabin/tree/leaseguard
} by modifying LogCabin, a key-value database atop the reference implementation of Raft (Section~\ref{sec:eval}). LogCabin lacked any lease mechanism until now.
Our experiments evaluate LeaseGuard’s impact on read and write latency, its availability after leader crashes, and how workload skewness affects read availability post-election. The results demonstrate that LeaseGuard achieves microsecond-latency local linearizable reads compared to many milliseconds of latency with LogCabin's default consistency mechanism, quorum checks. By avoiding quorum checks, LeaseGuard reduces the overhead of consistent reads from one to zero network roundtrips. It also improves write throughput from $\sim$1000 to $\sim$10,000 writes per second, by eliminating contention between writes and quorum reads. LeaseGuard improves read and write availability compared to traditional leases. Whereas traditional leases ban all reads on a new leader while it waits for a lease, in our LeaseGuard test the new leader instantly allows 99\% of reads to succeed.

\section{Background}
\label{sec:background}

\subsection{Raft}
\label{sec:raft}

\begin{figure}
    \centering
    \includegraphics[width=0.9\linewidth]{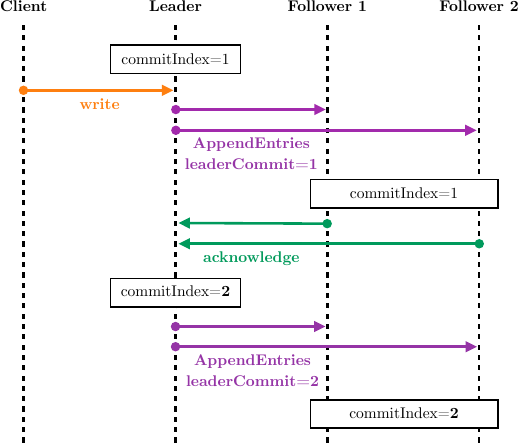}
    \caption{Raft replication. The leader advances its commitIndex when it learns that an entry is majority-replicated. Followers learn the new commitIndex from later AppendEntries messages.}
    \label{fig:Raft}
    \Description{A sequence diagram illustrating the Raft protocol. At first, the leader and both followers have a commit index of 1. A client sends a write to the leader, which sends an AppendEntries message to each follower. Once they acknowledge this message, the leader increments its commit index to 2, and tells the followers to increment their commit indexes to 2 also.}
\end{figure}

In Raft, each node has a \textit{state}: ``leader'', ``candidate'', or ``follower''. Each node has a \textit{term} that tracks the highest term number it has seen. Nodes gossip their term numbers during every communication.
To run for election, a follower increments its term and becomes a candidate, then requests votes from a majority of the replica set. Once elected, a node remains a leader until it crashes or observes another node with a higher term.

Clients invoke commands on the leader, which records them as \textit{entries} in its \textit{log} and sends them to followers in AppendEntries messages. Followers record entries in their own logs in the same order. An entry's \textit{log index} is its position in the log. Each node has a \textit{commitIndex}, the index of the latest entry it knows is durable. When the leader learns that a majority of nodes (including itself) have replicated up to a given log index, it advances its commitIndex to that point, i.e. it \textit{commits} the entry and all previous entries. It applies committed entries' commands to its local state machine and updates its \textit{lastApplied} index. The leader then replies to waiting clients, confirming their commands have succeeded. Followers eventually learn the leader's new commitIndex, which the leader sends them in subsequent AppendEntries messages. Thus followers' commitIndexes are less than or equal to the leader's in normal operation (Figure~\ref{fig:Raft}).

Raft and its predecessor MultiPaxos are similar~\cite{paxosVsRaft,ParallelsPaxosRaft}, but their election protocols differ significantly. In MultiPaxos, any node can be elected, and it must transfer from other nodes any log entries it lacks before it is fully functional. Raft, however, guarantees that a node is elected only if its log already includes all committed entries (``Leader Completeness''). Our work depends on this guarantee to provide novel simplifications and optimizations for leases.  

State Machine Replication protocols provide strongly-consistent, fault-tolerant write replication. But it is challenging to guarantee that reads conform to the strongest consistency level, linearizability~\cite{linearizability1990}. Linearizability requires that
(1) each operation appears to execute at a single instant between its invocation (when the client submits it) and completion (when the client receives the response), and 
(2) the execution order at these instants forms a valid sequential history, as if operations were executed atomically by a single thread.
As discussed in the introduction, a deposed leader in Raft may violate linearizability, if it serves a stale read assuming it is still the leader.

\subsection{Clocks with bounded uncertainty}
\label{sec:synchronized-clocks}

We assume each node has access to some function \textit{intervalNow()} which returns the interval [\textit{earliest}, \textit{latest}]. It is guaranteed that the true time was in this interval for at least a moment between the function's invocation and completion. Our lease algorithm requires a node to decide if a time recorded on another node is now more than $\Delta$ old, for some duration $\Delta$. For any two time intervals $t_1$ and $t_2$, a node knows that $t_1$ is more than $\Delta$ old if \textit{intervalNow()} has returned $t_2$ and $t_1.latest + \Delta<t_2.earliest$.

Google's Spanner paper~\cite{spanner} popularized the concept of bounded-uncertainty clocks in the cloud. It describes the TrueTime API, which returns an interval that is guaranteed to increase monotonically and to include the true time. Each Google data center has multiple time master machines; most have GPS receivers and the remainder have atomic clocks. Each Spanner server keeps its clock tightly synchronized, with known error bounds, by communicating with multiple time master machines and applying a variant of Marzullo's algorithm~\cite{marzullo}. TrueTime's average error was 4ms in 2012. Presumably its precision has improved, but it is still available only for Google's own systems, not Google Cloud Platform customers. Meta has deployed a similar infrastructure, also for their exclusive use, with sub-microsecond error~\cite{metaClocks}.

Starting in 2023, Amazon has deployed clocks with microsecond precision for public use. AWS TimeSync combines GPS receivers, the Precision Time Protocol~\cite{ptp}, and a physical hardware clock on each server to provide a high-precision time signal~\cite{aws-clocks}. Amazon's open source clock-bound~\cite{clock-bound} daemon monitors the local clock and calculates reliable error bounds. The AWS servers used in our evaluation have an average clock error less than 50\textmu s.

The Huygens protocol~\cite{huygens} and its commercialization, Clockwork~\cite{clockwork}, provide a similar guarantee, and they can be deployed on any cloud. As more cloud providers recognize the value of precise, bounded-uncertainty clocks, we hope and expect they will become more widely available.
\section{Adding leases to Raft}

\begin{figure}
\renewcommand\theFancyVerbLine{{\footnotesize\arabic{FancyVerbLine}}}
\begin{minted}[
fontsize=\small,escapeinside=@@,xleftmargin=20pt,linenos
]
{python}
# Handle a write request from a client.
def ClientWrite(command):
  if self.state != "leader": return "not leader"
  # Create new entry, log it and record its index.
  entry = (self.term, command, intervalNow())
  index = self.log.append(entry)
  # Another thread replicates, commits, and applies the
  # command, and advances lastApplied, see CommitEntry.
  await(self.lastApplied >= index)
  if self.state != "leader":
    # Deposed, don't know if command succeeded.
    return "not leader"
  return "ok"  

# Handle a read request from a client for key k.
def ClientRead(k):
  if self.state != "leader": return "not leader"
  # Last committed entry's age is calculated using
  # bounded-uncertainty clock.
  if self.log[self.commitIndex].age > delta:
    return "no lease"
  # Prevent "limbo" reads.
  if self.term != self.log[self.commitIndex].term:
    if @\texttt{any}@ limbo region entry affects k:
      return "key affected by limbo region"
  return self.data[k]

# When this node learns some followers have replicated
# entries up to index i, advance the commitIndex.
def CommitEntry(i):
  if self.state != "leader": return
  if not majorityAcknowledged(self.log[i]):
    return
  # Check for past-term entry < Delta old.
  # In reality this loop is optimized away, Sec. 7.
  for e in self.log:
    if e.term < self.term and e.age < delta:
      return
  self.commitIndex = max(self.commitIndex, i)
  while self.lastApplied<self.commitIndex:
    apply(self.log[self.lastApplied+1].command)
    self.lastApplied += 1
\end{minted}
\caption{LeaseGuard leader logic for a key-value store.}
\Description{A pseudocode listing}
\label{fig:leaseguard-logic}
\end{figure}

\label{sec:lease}

% Raft prevents a deposed leader from \textbf{committing writes} once a new (higher-term) leader has been elected. It also prevents a deposed leader from executing \textbf{reads}, but the Raft paper~\cite{raft} and thesis~\cite{ongaroThesis} propose either an expensive quorum-check per read, or a vaguely-specified lease protocol that may harm availability. LeaseGuard is a new mechanism with the same purpose: to prevent the deposed leader from executing reads while the new leader commits writes.

This section describes LeaseGuard using prose, pseudocode, and diagrams. A TLA+ model of LeaseGuard is available on GitHub\footnote{
https://github.com/muratdem/RaftLeaderLeases/tree/main/TLA
}.
We defer the correctness arguments to Section~\ref{sec:correct}. 

\textbf{Configuration:} All nodes have the same lease duration $\Delta$.

\textbf{Writes:} A leader can always accept a write command. It creates a log entry that includes the command, and replicates it to followers, as in Raft. In LeaseGuard, the leader includes its current time interval, \texttt{intervalNow()}, in the entry (Figure~\ref{fig:leaseguard-logic} lines 5-6).

\textbf{Advancing the commitIndex:} As in Raft, a leader can advance its commitIndex to \textit{i} once a majority of nodes have replicated the entry at \textit{i}. Once an entry is committed and applied, the leader acknowledges it to the client (lines 13 and 40-42). LeaseGuard adds a restriction: the leader $L_1$ of term \textit{t} cannot advance its commitIndex if $L_1$ has an entry < $\Delta$ old with term < \textit{t} from a deposed leader $L_0$ (lines 34-38). This prevents $L_1$ from committing writes while $L_0$ may have an active lease and may be executing reads, see Figure \ref{fig:LeaseTransition}.

\textbf{Reads:} A leader \textit{L} can serve a local linearizable read if it has a committed entry < $\Delta$ old in any term, because it knows no newer leader is committing writes (lines 18-21). If \textit{L} has not yet committed an entry in its term, there is a ``limbo region'' of entries past \textit{L}'s commitIndex, up to and including the last entry \textit{L} had when it was elected. \textit{L} does not yet know if an earlier leader committed any of these entries, so \textit{L} executes a read only if its result is not affected by any entry in the limbo region (lines 23-25).

\textbf{Followers:} LeaseGuard does not change follower behavior compared to vanilla Raft. Only when a follower is elected as a leader does it make use of the lease information implied by its log. 

\textbf{Elections:} In contrast to previous work~\cite{spanner,crdb,yugabyteLeases,Liskov2012}, we leave Raft's election protocol~\cite{raft} unmodified. Since our priority is to maximize availability, we allow a new leader to be elected before the old leader's lease expires. Even a node that knows of a valid lease may vote, become a candidate, and/or become a leader. 

Figure~\ref{fig:leaseguard-logic} presents the pseudocode for LeaseGuard. The following sections focus on certain details of LeaseGuard.

\begin{figure}
    \centering
    \includegraphics[width=1\linewidth]{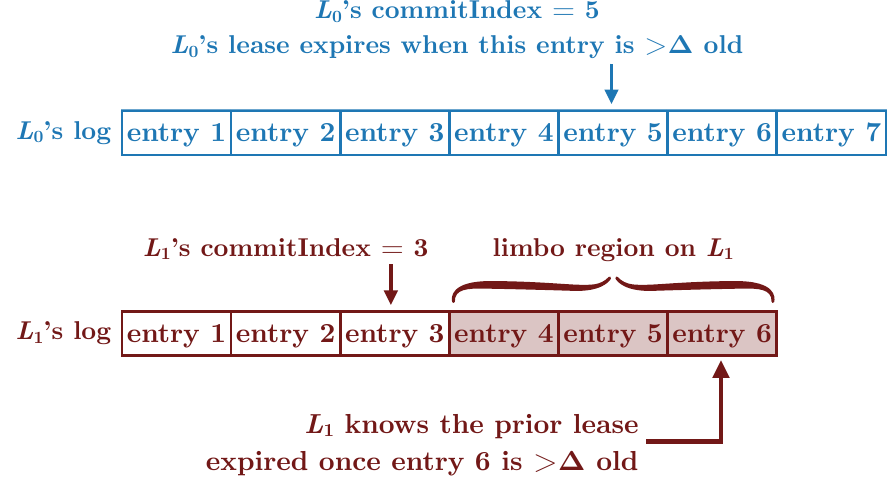}
    \caption{Logs of old leader $L_0$ and new leader $L_1$ just after $L_1$ was elected, before it commits any entry in its term.}
    \label{fig:limbo-range}
    \Description{A diagram illustrating which log entries are in the limbo region. There are two leaders L0 and L1, each with a log. L1 was elected after L0, and it hasn't committed any entries in its term. L0's log has 7 entries and its commit index is 5. L1's log has 6 entries and its commit index is 3. L1 has a limbo region spanning from 4 (one past its commit index) to 6 (the end of its log when it was elected).}
\end{figure}

\subsection{Establishing a lease}
Leases are naturally established and extended by Raft's replication protocol. A leader $L_1$ establishes a lease on followers by creating an entry \textit{e} in its log and sending it to followers. 
When $L_1$ commits \textit{e} (after hearing acks from a majority of nodes), $L_1$ can serve local linearizable reads while \textit{e} is at most $\Delta$ old, because it knows no future leader will advance the commitIndex until \textit{e} is more than $\Delta$ old. It knows this because, once \textit{e} is committed, Leader Completeness implies that any leader $L_2$ in a future term will have \textit{e} in its log, and thus know of $L_1$'s lease. Later entries after \textit{e}, including ordinary client write commands, automatically extend the lease. 

Expressed another way, once a leader has committed \textit{any} entries, its newest committed entry is its lease. On the other hand, a new leader that \textit{hasn't} committed entries considers its newest (possibly uncommitted) entry from the prior term as the \textit{prior} leader's lease.

\subsection{Deferred commit writes}
\label{sec:deferred-commit-writes}

In existing protocols, a non-leaseholder cannot accept writes. In LeaseGuard a leader can always accept writes, send them to followers, and make them fault-tolerantly durable. It simply cannot commit, apply, or acknowledge writes until its latest prior-term entry (i.e., the deposed leader's lease) is more than $\Delta$ old. We call this the \emph{deferred commit optimization}. When the prior-term entry is old enough, the leader advances its commitIndex to the index of the latest majority-replicated entry, and acknowledges all the pending writes that are now committed.

This optimization lets the leader prepare writes to be committed as soon as the old lease expires without requiring further follower communication. Even if the leader crashes or is deposed before these writes are committed, they are durable once majority-replicated, and will be committed by a future leader. By accepting writes while waiting for a lease, the new leader avoids being overwhelmed by a thundering herd of writes the moment its lease begins.

\begin{figure}
    \centering
    \includegraphics[width=1\linewidth]{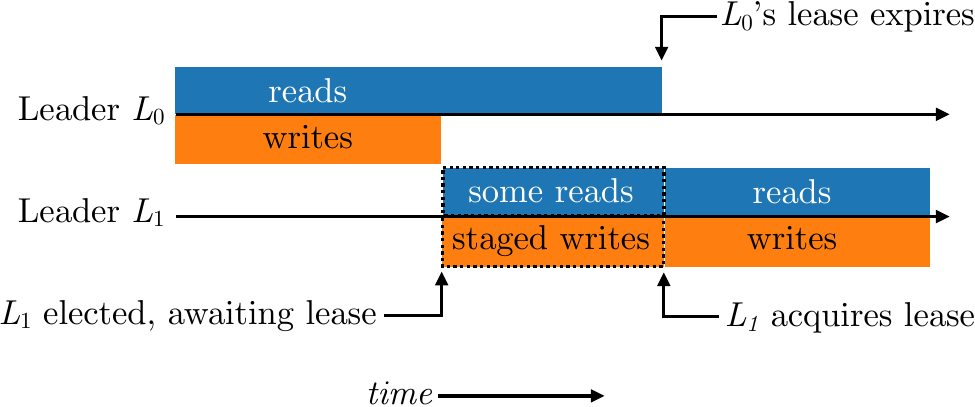}
    \caption{Transitions in the read/write availability of leaders with LeaseGuard. While the new leader waits for a lease, it can serve some consistent reads and accept writes. Meanwhile the old leader serves reads.}
    \label{fig:LeaseTransition}
    \Description{A flowchart showing when leaders can read or write with leases. There are two leaders. At first, Leader 1 can execute reads and writes. Then Leader 2 is elected. Now Leader 1 can execute only reads, while Leader 2 can execute reads unaffected by the limbo region, and it can stage writes. Once Leader 1's lease expires, Leader 2 can execute reads and writes freely.}
\end{figure}

\subsection{Inherited lease reads}
\label{sec:inherited-lease-reads}

LeaseGuard allows multiple leaders to read simultaneously, significantly enhancing read availability during leadership transitions (Figure~\ref{fig:LeaseTransition}). Consider a leader $L_0$ that has been deposed by $L_1$. $L_0$ may still believe it is the leader and can serve reads while its last committed entry $e$ is < $\Delta$ old. $L_1$ (and any future leader) is guaranteed to have $e$ (Leader Completeness) and will not advance its own commitIndex until $e$ is more than $\Delta$ old. Thus $L_0$ has the latest committed data in the replica set while its lease is valid.

In LeaseGuard, we propose that $L_1$ can also serve linearizable reads with the lease it inherited from $L_0$. $L_1$ has all the entries $L_0$ committed. It knows $L_0$ cannot commit any more entries in its leadership term, because when $L_1$ won the election it increased a majority of the nodes' terms and prevented $L_0$ from committing any more entries~\cite{raft}. Since $L_1$ has the latest committed data, this might seem to guarantee that $L_1$ can execute linearizable reads, but there is an additional kink: we need to take the ``limbo region'' of the log into account.

Consider Figure \ref{fig:limbo-range}. $L_1$ has been elected in a later term than $L_0$, but $L_1$ has not yet committed any entries, and its commitIndex lags that of $L_0$. We define $L_1$'s ``limbo region'' as entries in $L_1$'s log with indexes greater than $L_1$'s commitIndex, up to the last entry in $L_1$'s log when it was elected. $L_1$ does not know where in the range [3,6] $L_0$'s commitIndex falls.
If $L_1$ acts pessimistically and executes client reads from its commitIndex, it may return older data than $L_0$ (if $L_0$ is alive), violating linearizability. If $L_1$ acts optimistically and applies all commands through index 6, it may return newer data than $L_0$, which also violates linearizability. 

Therefore, LeaseGuard adds an additional check: a node executes reads \textbf{only if they are unaffected by writes in the limbo region}. This way both $L_0$ and $L_1$ can serve linearizable reads. Once $L_0$'s lease expires, $L_1$ commits an entry and the limbo region disappears. For a simple key-value store, it is self-evident whether a write in the limbo region affects some point query or range query. Databases that support more complex queries require specialized algorithms. For example, if a database supports multi-version concurrency control (MVCC), the leader could execute a query on each version of its data corresponding to each entry in the limbo region, and reject the query if any results are unequal. To make this practical is the subject of future research.

\section{Correctness}
\label{sec:correct}

In this section, we show that LeaseGuard guarantees linearizability. Section~\ref{sec:raftGuarantees} describes Raft's guarantees. Section~\ref{sec:correctBasic} shows that LeaseGuard is correct, assuming perfect clocks. We consider clock uncertainty in Section~\ref{sec:correctSkew}, and discuss why correctness is preserved under reconfiguration in Section~\ref{sec:correctRecon}.

\subsection{Raft guarantees}
\label{sec:raftGuarantees}

We rely on the following guarantees, which we quote from~\cite{raft}:

\noindent
\textbf{Leader Append-Only}: a leader never overwrites or deletes entries in its log; it only appends new entries.

\noindent
\textbf{Log Matching}: if two logs contain an entry with the same index and term, then the logs are identical in all entries up through the given index.

\noindent
\textbf{Leader Completeness}: if a log entry is committed in a given term, then that entry will be present in the logs of the leaders for all higher-numbered terms.

\noindent
\textbf{State Machine Safety}: if a server has applied a log entry at a given index to its state machine, no other server will ever apply a different log entry for the same index.

\subsection{Correctness with perfect clocks}
\label{sec:correctBasic}

A concurrent computation is linearizable if it is equivalent to a sequential computation that preserves the real-time ordering of operations; that is, each operation appears to take effect at some instant between its invocation and acknowledgment~\cite{linearizability1990}. Raft without LeaseGuard permits multiple leaders at a time, but it is nevertheless linearizable. Raft guarantees that only the highest-term leader can commit writes, that it commits and applies each write between its invocation and acknowledgment, and that writes are applied sequentially in the same order on all nodes. To ensure linearizable reads, Raft performs a quorum check for each read to ensure the leader executing the read is the highest-term leader.

LeaseGuard alters Raft's behavior in two ways. First, a leader may wait before committing recent writes and acknowledging them to the client (\textit{deferred commit write}, Section~\ref{sec:deferred-commit-writes}). Raft has no real-time bound on when a leader commits an entry after it is majority-acknowledged, so this delay cannot break Raft's linearizability guarantee.

Second, a leader can execute a read without contacting other nodes (\textit{inherited lease reads}, Section~\ref{sec:inherited-lease-reads}). We sketch a proof that LeaseGuard guarantees Read-Your-Writes, and therefore guarantees linearizability. 

\begin{thm}[Read-Your-Writes]\label{thm:ReadYourWrites} 

If a leader $L_1$ in term $t_1$ commits and applies write $w$, represented by log entry $e_1$ with index $i_1$, a later read $r$ must observe $w$'s effect.
\end{thm}

To prove this, we show that $r$ observes $w$'s effect in three cases:

\textbf{Case 1:} $r$ is sent to $L_1$ while it is still the leader in term $t_1$. We know that $w$ was applied on $L_1$, and Raft prevents undoing writes, so Theorem~\ref{thm:ReadYourWrites} holds. If $L_1$ becomes a leader again in a future term, we call it $L_2$ and handle it in Case 3.

\textbf{Case 2:} $r$ is sent to a deposed leader $L_0$ with term $t_0<t_1$. $L_1$ has $L_0$'s last committed entry $e_0$ in its log by Leader Completeness. LeaseGuard requires $L_1$ to wait until all prior-term entries in its log are more than $\Delta$ old before committing an entry in $t_1$. Since $L_1$ has committed $e_1$, we know that $e_0$ is more than $\Delta$ old. LeaseGuard requires $L_0$ to have a committed entry less than $\Delta$ old in order to read (Section~\ref{sec:inherited-lease-reads}), but $e_0$ is $L_0$'s last committed entry and it is older than $\Delta$; thus $L_0$ rejects $r$ and upholds Theorem~\ref{thm:ReadYourWrites}.

%\vspace{3mm}
\textbf{Case 3:} $r$ is sent to leader $L_2$ with term $t_2>t_1$. We subdivide this case according to the state of $L_2$'s log and commitIndex.

%\vspace{3mm}
\textbf{Case 3.1:} $L_2$'s commitIndex is $i_0 \leq i_1$ and $L_2$ has no limbo region. 
%(uncommitted prior-term entries in $L_2$'s log, Section~\ref{sec:inherited-lease-reads}).
By Leader Completeness, $e_1$ is in $L_2$'s log, thus since it has no limbo region $i_0 = i_1$. By State Machine Safety, $L_2$ has applied or will eventually apply $w$. Raft requires $L_2$ to wait until its lastApplied reaches the commitIndex $L_2$ had when $r$ arrived, before executing $r$ (see ClientRead in Figure~\ref{fig:leaseguard-logic}). So $L_2$ applies $w$ before executing $r$, and satisfies Theorem~\ref{thm:ReadYourWrites}.

%\vspace{3mm}
\textbf{Case 3.2:} $L_2$'s commitIndex is $i_0 \leq i_1$ and it has a limbo region, meaning that it has not committed an entry in its term. $L_2$'s limbo region spans from its commitIndex $i_0$ to its last prior-term log index $i_2$, where $i_0 \le i_1 \le i_2$, so the limbo region includes $w$. According to the limbo read rule in Section \ref{sec:inherited-lease-reads}, $L_2$ will not execute $r$ if it is affected by a command in the limbo region. $L_2$ will reject $r$ if it is affected by $w$, otherwise execute $r$; either way the client cannot observe that $w$ has not been executed, so Theorem~\ref{thm:ReadYourWrites} is upheld. 

%\vspace{3mm}
\textbf{Case 3.3:} $L_2$'s commitIndex is $i_2 > i_1$. Thus it has no limbo region, and it must eventually apply $w$ (State Machine Safety). It must apply $w$ some time before it executes $r$, 
satisfying Theorem~\ref{thm:ReadYourWrites}.
% It will execute $r$ when its lastApplied equals its commitIndex, 

Theorem~\ref{thm:ReadYourWrites} holds in all cases: read $r$ observes write $w$'s effect regardless of whether $r$ is sent to the leader that committed $w$, an earlier-term leader, or a later-term leader (with or without a limbo region or a committed entry in its term).

Now we show that Theorem~\ref{thm:ReadYourWrites} implies that reads reflect the latest committed state among all leaders in the replica set. Note that it is theoretically possible for a replica set with $2f+1$ nodes to have $f+1$ leaders simultaneously: each received $f+1$ votes from itself and its $f$ non-leader peers. In real systems, more than two leaders is practically unheard of.

\begin{thm}[Read At Highest commitIndex]\label{thm:commitIndex}
For the set of nodes $\mathcal{L}$ whose state is ``leader'' at some moment, let $i_{max}$ be the greatest commitIndex over $\mathcal{L}$. Any read $r$ observes the effects of all commands $c_i$ associated with all entries $e_i$ that were ever committed by any leader, for all $i \leq i_{max}$, for the value of $i_{max}$ at some moment between $r$'s invocation and acknowledgment.
\end{thm}

Proof: Suppose $r$ observes some earlier version of the state machine before the entry with index $i_{max}$ was applied. There must be a write $w$ associated with the $i_{max}$ entry, whose effect $r$ does not observe. This contradicts Theorem~\ref{thm:ReadYourWrites}, thus $r$ cannot observe the state machine at an earlier version.

\vspace{3mm}

Raft guarantees that the sequence of state machine versions at sequential commitIndexes is a linearizable history, and that the latest commitIndex in the replica set increases monotonically. Since LeaseGuard ensures that each read observes the latest commitIndex in the replica set, LeaseGuard preserves linearizability.

\subsection{Correctness with clock uncertainty}
\label{sec:correctSkew}

In Case 2 above, a node must determine whether a time interval was recorded on another node more than $\Delta$ ago. This requires clocks with correctly measured uncertainty bounds on each node, such that $t_1$ is more than $\Delta$ old if \textit{intervalNow()} returns $t_2$ and $t_1.latest + \Delta<t_2.earliest$. Such clocks have become available in the cloud over the last two years (Section~\ref{sec:synchronized-clocks}). If a clock's bounds are wrong, i.e. if the true time is outside the interval [\textit{earliest}, \textit{latest}] for the entire duration between the invocation and completion of a call to \textit{intervalNow()}, then multiple leaders may think they are leaseholders at once. In that case a deposed leader may execute reads while a higher-term leader commits writes. Reads from the deposed leader will violate Read-Your-Writes, and thus violate linearizability. Inherited lease reads require correct clock bounds!

\subsection{Correctness with reconfigurations}
\label{sec:correctRecon}

Raft has two protocols for ``reconfiguration'', i.e. adding and removing replica set nodes: joint consensus, or single-node changes~\cite{loglessRecon}. Both protocols uphold all the Raft guarantees we rely on, including Leader Completeness and State Machine Safety. With \textbf{joint consensus}, a replica set with configuration $C_{\text{old}}$ switches to configuration $C_{\text{new}}$ by adding and removing arbitrary members. The replica set transitions through a period when all entries must be committed by a majority of both $C_{\text{old}}$ and $C_{\text{new}}$, then it retires $C_{\text{old}}$ and makes all subsequent decisions with a majority of $C_{\text{new}}$ only. With \textbf{single-node changes}, the replica set can either add one node or remove one node per reconfiguration, and there is no transitional period. The protocol is safe because a majority of the old and new configuration always overlaps if they differ by one server. This overlapping-majorities property preserves the Raft guarantees on which LeaseGuard depends.
\section{Extensions and optimizations}
\label{sec:opt}

\subsection{Automatic lease extension}
\label{sec:automatic-lease-extension}
LeaseGuard extends leases automatically through write operations, but a lease expires after $\Delta$ duration with no writes. When writes are rare, the leader can write a no-op to reestablish its lease whenever needed to serve a read, or (to avoid cold starts) proactively write a no-op to maintain its lease. It is also possible to change $\Delta$ dynamically, e.g. to increase $\Delta$ to reduce the frequency of no-ops. This is out of our current scope.
For planned leader transitions (e.g. for rolling upgrades and other maintenance tasks), the outgoing leader can relinquish its lease by committing an ``end-lease'' entry as its final act of leadership. The next leader can execute reads and writes without restriction.

\subsection{Choosing election timeout and $\Delta$}
\label{sec:choice-of-lease-duration}

\begin{figure}
    \centering
    \includegraphics[width=1\linewidth]{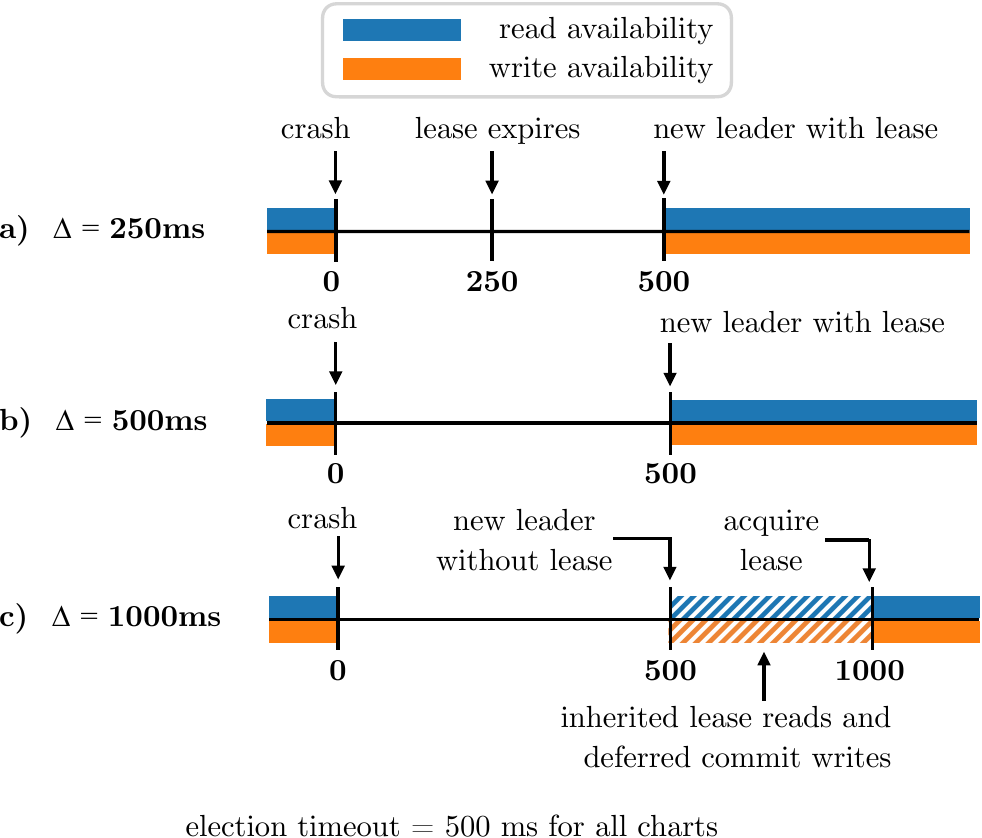}
    \caption{Effect of lease duration on availability in LeaseGuard. Election timeout = $\Delta$ is usually optimal.}
    \label{fig:availability-timeline}
    \Description{A timeline showing how three different lease durations will affect availability after an election. The election timeout is 500 ms in all scenarios. In scenario A, the lease duration is 250 ms. The current leader crashes at time 0, its lease expires at time 250, and a new leader with a lease is elected at time 500. Since the lease duration was shorter than the election timeout it had no effect on availability. In scenario B, the lease duration is 500 ms, the same as the election timeout. As before, the old leader crashes at time 0, and a new leader with a lease is elected at time 500. In scenario C, the lease duration is 750 ms, which is longer than the election timeout. In this scenario a leader is again elected at time 500, but it has no lease until time 750, so it spends 250 ms without a lease. During this time it can stage deferred commit writes and execute inherited lease reads.}
\end{figure}

LeaseGuard allows the election timeout $\Etime$ and lease duration $\Delta$ to be tuned independently. If $\Etime$ is a fixed number, it is best to set $\Delta=\Etime$ (Figure~\ref{fig:availability-timeline}). 
If $\Delta<\Etime$ and writes are rare, the leader must extend its lease more often without any benefit from the short $\Delta$: the long election timeout causes unavailability after a crash, regardless of leases. 
If $\Delta>\Etime$, there is a period after an election when the new leader has no lease. 
LeaseGuard tolerates this better than prior protocols, because the elected leader can still serve linearizable reads using the inherited lease optimization and perform deferred commit writes (see evaluations in Sections~\ref{sec:simulation} and~\ref{sec:eval}). 
Regardless of $\Etime$, a leader might be deposed at any time, perhaps by a human operator or a failure detector. We anticipate that LeaseGuard will enable deployment of more aggressive failure detectors, since it improves availability after an election.

\subsection{Leases without bounded-uncertainty clocks}
\label{sec:timer}
Most of LeaseGuard can be implemented using local \textit{timers} with bounded drift rates; bounded-uncertainty clocks are not required. All nodes must know some value $\epsilon$, the maximum amount that their clocks can gain or lose while measuring a duration of $\Delta$. Whenever a leader creates or a follower replicates an entry, it starts a timer to track that entry's age. The leader's timer starts as soon as it creates an entry, before any followers can replicate the entry, so the leader's timer expires before any follower's timer for that entry. LeaseGuard's rules can be revised to use timers instead of time intervals: a leader can advance its commitIndex or serve reads so long as its last entry in any previous term is > $\Delta + \epsilon$ old, and the last committed entry in its own term is < $\Delta - \epsilon$ old.

Timers suffice for LeaseGuard with deferred commit writes, but not for inherited lease reads: Suppose a leader $L_1$ is deposed by $L_2$, which commits no entries before it is deposed by $L_3$. Due to a network partition, $L_2$ is unaware of $L_3$'s election; $L_3$ is elected by a quorum that does not include $L_2$. Now $L_2$ and $L_3$ are both leaders at the same time. They may disagree on when $L_1$'s lease expires, based on when they last replicated an $L_1$ entry when they were followers.
This disagreement can lead to a linearizability violation: $L_3$ believes $L_1$'s lease has expired, so $L_3$ commits entries, while $L_2$ thinks the lease it inherited from $L_1$ is still valid, so $L_2$ serves inherited lease reads from stale data, which does not reflect $L_3$'s writes. Thus, systems must have access to clocks with bounded uncertainty to implement inherited lease reads. With such clocks, $L_1$ stores a time interval in each entry, and there is no ambiguity about whether its entries are > $\Delta$ old.

\subsection{Lessons learned}

Our experience offers several lessons for the design of replicated databases. First, we found that Raft's {\em Leader Completeness} property provides a surprisingly powerful foundation for reasoning about leases: by tying lease validity to log replication rather than to ad-hoc timers or heartbeats, we eliminate entire classes of gray-failure and faux-leader bugs that plagued production systems. 

Second, our work reinforced the value of formal methods not only as a tool for verification, but also as a tool for design discovery: both our deferred commit and inherited lease optimizations emerged from exploring corner cases in our TLA+ model. At the same time, our early iterations reminded us that formal tools are not foolproof. Our first specification missed the ``limbo-region'' read problem because we modeled using large atomic actions that hid the bug. This experience taught us to approach even successful model checks with skepticism: as Lamport puts it, ``always be suspicious of success.'' 

Finally, reasoning through these bugs and correctness arguments led us to view LeaseGuard through the lens of {\em knowledge and common knowledge in distributed systems}~\cite{halpern1990knowledge}. We are exploring how to make this epistemic-reasoning approach more practical for designing distributed database protocols.

\section{Simulation}
\label{sec:simulation}

We wrote a Raft simulation in Python. Simulation is helpful for quickly exploring performance characteristics, which are impossible to analyze with a TLA+ specification, and difficult to measure cleanly with a real system~\cite{formalMethodsSolveHalfMyProblems,simulationBrooker,cheng_barbarians_2025}. Our simulation is at a level of abstraction somewhere between our specification and our implementation. It includes I/O latency, periodic no-ops for lease extension, and other details we omit in TLA+. On the other hand, it is far more abstract than the C++ implementation: it has no storage layer, network protocol, or query language. The simulation runs in a single process. It represents nodes as Python objects that pass messages. Time, clock error bounds, thread scheduling, and network delays are all simulated. For nondeterministic events such as checking an imperfect clock, or sending a message with unpredictable latency, we choose appropriate probability distributions and produce values with a pseudorandom number generator (PRNG). For a given seed and set of parameters, the PRNG produces the same sequence of values, thus the simulator executes the same events. We carefully engineered this reproducibility to ease debugging and analysis. The simulator expresses the LeaseGuard algorithm more simply and directly than the implementation, providing a useful reference (in addition to the TLA+ spec and C++ code) for implementers.
We implemented two consistency mechanisms in our Raft simulator: quorum checks as described in the Raft paper, and LeaseGuard.

\subsection{Simulator architecture}
The simulator comprises these files (about 1000 lines of code, excluding whitespace and comments):

\vspace*{2mm}
\noindent
\texttt{simulate.py}: An event loop. Callbacks are scheduled to run at times in the future. The event loop finds the callback with the earliest deadline, advances the simulated clock to exactly that deadline, executes the callback (which may schedule more callbacks), then repeats. Atop the loop we layered an API with tasks, futures, and coroutines, similar to Python's standard \texttt{asyncio}. Coroutines simulate processes and threads, but they are deterministic, with the same scheduling and interleaving each time the simulation runs with a given seed.

\vspace*{2mm}
\noindent
\texttt{lease\_guard.py}: Implements Raft with various consistency mechanisms, including LeaseGuard. Any number of nodes can form a replica set (our experiments use three nodes). Nodes communicate via message-passing, with random network delays. Each node has an independent, bounded-uncertainty clock. Nodes expose two commands to clients: \texttt{write(key, value)} permits a client to append \texttt{value} to the append-only list associated with \texttt{key}, and \texttt{read(key)} returns the values appended to this list, in order. We use append-only lists because they are ideal for checking that our algorithm upholds linearizability~\cite{elle} (Section~\ref{sec:linearizability-checking}).

\vspace*{2mm}
\noindent
\texttt{client.py}: Simulates a client. Any number of clients can concurrently communicate with the Raft leader. Each client appends one unique value to one key, or reads the list of values for one key. For our experiments we implement workload generators that launch many concurrent clients. These workload generators are \textit{open loop}: they start requests at a fixed rate regardless of the response latency.

\vspace*{2mm}
\noindent
\texttt{params.py}: Contains parameters that control every aspect of the Raft and LeaseGuard protocols, e.g. whether leases are enabled, which optimizations are enabled, the election timeout, and lease duration. The module also includes parameters that control the simulation, e.g. maximum clock error bounds, the network latency mean and variance, the number of clients and their arrival rate.

\vspace*{2mm}
\noindent
\texttt{prob.py}: Produces pseudorandom numbers with various probability distributions.

\vspace*{2mm}
\noindent
\texttt{run\_with\_params.py}: Reads parameters from \texttt{params.py}, configures a replica set, waits for it to elect a leader, concurrently runs a large number of clients, gathers statistics such as average read/write latency, and checks that the execution was linearizable.

\subsection{Linearizability checking}
\label{sec:linearizability-checking}

We implemented a linearizability checker in Python to catch bugs in \texttt{lease\_raft.py}, i.e. nonconformance with the TLA+ specification. 
% See Appendix \ref{sec:appendix-a} for the linearizability-checking code. 
Each run of the simulator compiles a history of operations. A history is a sequence of \texttt{ClientLogEntry} objects with these fields:

\begin{itemize}
    \item \texttt{op\_type}: ListAppend or Read.
    \item \texttt{start\_ts}: True time the client started the operation.
    \item \texttt{execution\_ts}: True time the operation completed.
    \item \texttt{end\_ts}: True time the client received the reply.
    \item \texttt{key}: The key that was read from or written to.
    \item \texttt{value}: The appended value for ListAppend, or the returned list of values for Read.
    \item \texttt{success}: Whether the operation succeeded.
\end{itemize}

Linearizability checking in general is NP-complete~\cite{gibbons_99} because it requires evaluating all serializations that respect real-time order. Our simulator has the advantage of omniscience: it knows the true time when all client and server events occur. Each ListAppend entry’s \texttt{execution\_ts} records when the write was committed on the leader, and each Read entry’s \texttt{execution\_ts} marks when it was executed. To check linearizability, we check that each operation's execution time falls between its start and end times, then sort operations by execution time, and verify that Reads observe all preceding ListAppends for the same key. In some cases we still must consider multiple serializations: (1) If multiple ListAppends share the same exact execution time, we test all orderings. The history is linearizable if at least one ordering preserves ListAppend and Read semantics. (2) If a ListAppend fails from the client’s perspective, it may have succeeded on the server, e.g., if the leader was deposed after replication but before responding. We check linearizability under both possibilities: the ListAppend either succeeded after it started or never succeeded. Despite these permutations, checking simulated histories with hundreds of operations usually completes in a few seconds.

\subsection{Simulated experiment design}
\label{sec:experiment-design-simulation}
We perform simulated experiments to evaluate the following:

\noindent
{\bf Q1} (Latency): How do several consistency mechanisms affect the latency of linearizable reads and writes?

\noindent
{\bf Q2} (Availability): How do several consistency mechanisms affect availability after a leader crashes?

\noindent
{\bf Q3} (Skewness): How does workload skewness affect read availability after a leader crashes?

\subsection{Latency simulation}
\label{sec:latency-simulation}

\begin{figure}
    \centering
    \includegraphics[width=1\linewidth]{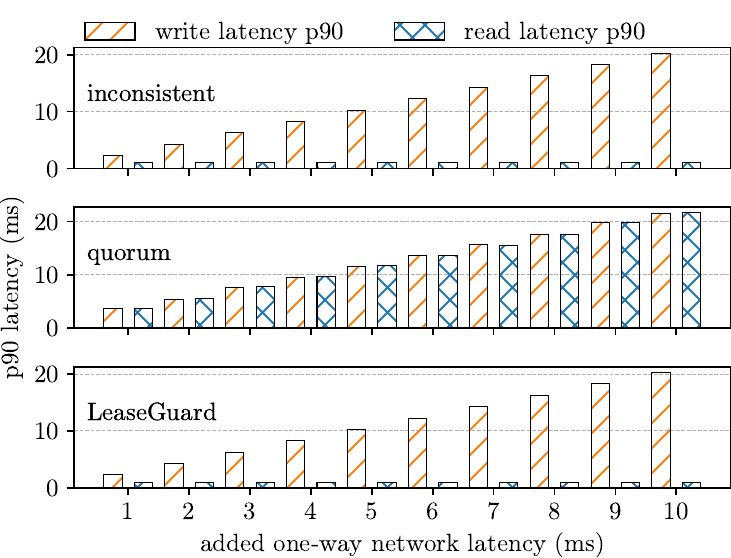}
    \caption{Effect of network latency on read/write latency, milliseconds (simulation). The quorum check mechanism makes reads as slow as writes. LeaseGuard makes consistent reads as fast as inconsistent reads.}
    \label{fig:network-latency}
    \Description{Three bar charts showing the effect of network latency on read/write latency, depending on what consistency mechanism we use. If we allow inconsistent reads, then reads are always instantaneous in our simulation, and write latency is about twice network latency. If we use quorum reads, then read latency is about twice network latency, the same as reads. With LeaseGuard, reads are instantaneous, and write latency is twice network latency as usual.}
\end{figure}

For Q1, we simulate a range of one-way network latencies between server nodes, to explore the effects of latencies within and across regions. We use a lognormal distribution with means from 1-10ms and with variance equal to the mean, to simulate the latencies we would observe within and across regions. Client-server latency is zero. There are 50 clients, half do a single Read operation and half do a single ListAppend. Clients arrive according to a Poisson process, with an average of 100ms between arrivals.

Figure~\ref{fig:network-latency} shows how network latency affects read and write 90th-percentile latency in three configurations. The \textbf{inconsistent} configuration does not ensure Read-Your-Writes consistency during elections. Write latency is determined by the time it takes the leader to replicate the write to at least one follower and receive acknowledgment. Read latency is zero in our simulation, since the leader serves reads without communicating with followers. The \textbf{quorum} configuration uses Raft's default consistency mechanism: the leader exchanges messages with a majority of nodes for each read, therefore read latency is determined by the round-trip time between the leader and at least one follower. The 90th-percentile write latency is higher in this configuration (6\% higher with 10ms one-way network latency), because replication competes with quorum checks for I/O. The \textbf{LeaseGuard} configuration uses our protocol, which enables instant linearizable reads served locally. Writes are as fast as the ``inconsistent'' configuration, since replication no longer competes with reads for I/O. LeaseGuard improves consistency with no overhead.

\subsection{Availability simulation}
\label{sec:availability-simulation}

\begin{figure}
    \centering
    \includegraphics[width=1\linewidth]{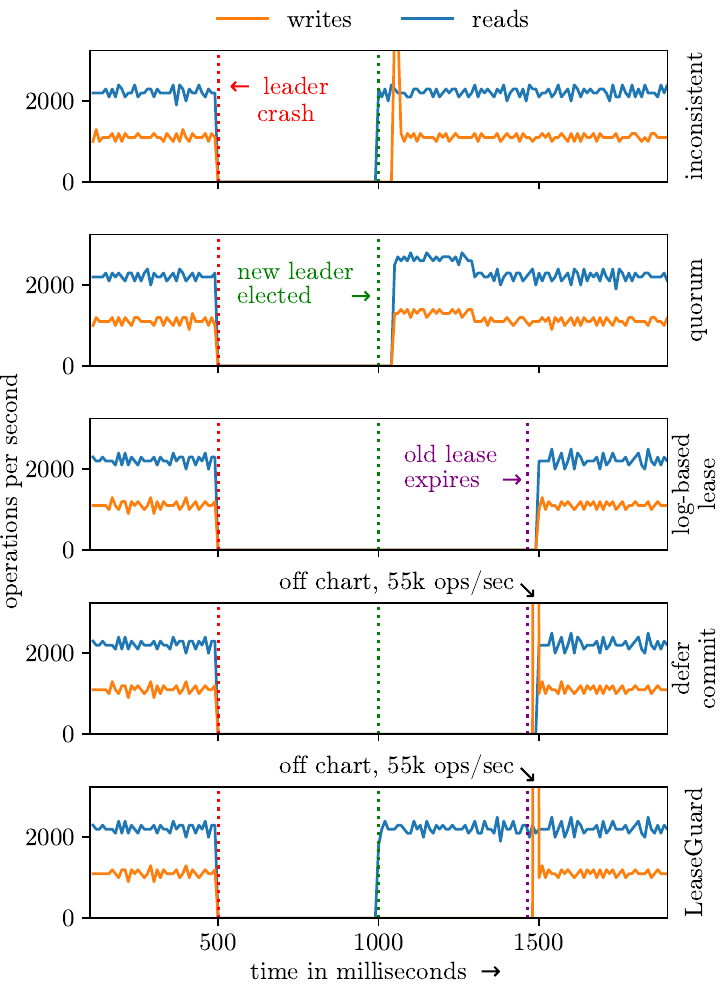}
    \caption{Availability (simulation). $\Delta =$ 1 second, $\Etime =$ 500 ms. Just after failover, LeaseGuard's deferred commit optimization increases write throughput, and its inherited lease reads improve read availability.}
    \label{fig:availability}
    \Description{Five stacked line charts, with a shared horizontal time axis measured in milliseconds since the start of the experiment. The five configurations and their availability characteristics are described in the text.}
\end{figure}

For Q2, we used network latency parameters derived from AWS same-subnet statistics (191 $\mu$s mean, 391 $\mu$s variance)~\cite{hilyard_cloudy_2023}. An open-loop workload generator starts an operation every 300\textmu s. One-third of operations write 1 kilobyte to a random key, two-thirds read a random key. Keys are uniformly randomly selected from a list of 1000. The election timeout $\Etime$ is 500ms. (This value is within the range used by other Raft implementations; \cite{raft} proposes timeouts from 12ms to 300ms. Production systems use larger timeouts, typically 1 to 10 seconds to avoid spurious elections, e.g. CockroachDB's default is 9 seconds and TiDB Placement Driver's is 10.) We set lease duration $\Delta$ to 1 second (twice $\Etime$) to observe leader-transition availability effects. In practice, a system that uses a constant $\Etime$ to detect leader failures should set $\Delta=\Etime$, see Section~\ref{sec:choice-of-lease-duration}. However, a sophisticated system might use a dynamic failure detector that calls for an election whenever it suspects the leader, even when its lease is a long time from expiring. Setting $\Delta=2\Etime$ is a simple way to examine this scenario. We study the effects of various consistency mechanisms on availability (Figure~\ref{fig:availability}). In each experiment we create a 3-node replica set and begin executing the workload. After 500ms, the leader crashes; 500ms later another leader is elected, and 500ms after that the old leader's lease expires.

\textbf{Inconsistent:} No mechanism ensures Read-Your-Writes. After the first leader crashes, all reads fail. Once the new leader is elected, they resume at their normal throughput instantly. Writes also fail during the interregnum between the old and new leader. After the election, clients resume writing to the new leader, but the leader has not yet committed any writes. There is a short delay while the surviving follower discovers the new leader and begins replicating. Write throughput then appears to spike as the follower catches up and acknowledges recent log entries, and the leader commits them and acknowledges them to the client. This option has good availability but does not guarantee linearizability.

\textbf{Quorum:} The leader communicates with a majority of nodes (itself and at least one follower) for each read/write. All operations fail during the interregnum, and reads and writes are both delayed a short time after the election. Since reads and writes compete for I/O, they both show a throughput spike after the election, but the spike is lower than the write spike in the ``inconsistent'' scenario. Once the new leader has cleared its replication backlog, throughput returns to the steady state. This option has good availability, at the cost of a lower throughput ceiling and higher latency (Section~\ref{sec:latency-simulation}).

\textbf{Log-based lease:} This is Raft with LeaseGuard, but without inherited read leases or deferred-commit writes. After the election, reads and writes fail until 1500ms when the new leader acquires its lease. (An implementer could choose to retry operations instead of fail-fast.)

\textbf{Defer commit:} The new leader begins to accept writes, create log entries, and replicate them as soon as it is elected, while waiting for the old lease to expire. However, it does not advance the commit index until it acquires a lease at 1500ms. These writes are all acknowledged near-simultaneously when the leader fast-forwards its commit index, so write throughput goes off the chart and the replica set returns to steady state quickly. The deferred commit write optimization allows the system to stabilize more quickly than with unoptimized leases.

\textbf{LeaseGuard:} Our lease protocol with all optimizations enabled. The new leader can serve linearizable reads as soon as it is elected, without waiting for the old lease to expire. There are few writes in progress when the leader dies (since write interarrival times are large compared to network latency), so the new leader has a small limbo region which does not significantly interfere with reads. The inherited lease read optimization restores read availability sooner after an election.

In summary: When $\Delta>\Etime$, leader leases inevitably hurt availability. But this experiment shows that LeaseGuard's two optimizations, inherited read leases and deferred-commit writes, help the replica set return to steady state more quickly after a leader failure.

\subsection{Skewness simulation}
\label{sec:skewness}
\begin{figure}
    \centering
    \includegraphics[width=1\linewidth]{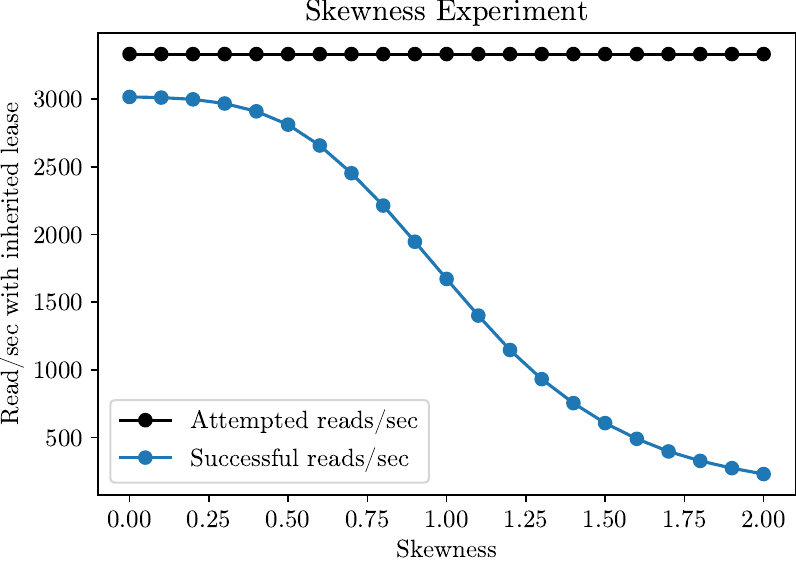}
    \caption{Effect of workload skewness on read throughput on the new leader while it awaits a lease (simulation). More skew means fewer consistent reads are permitted. Throughput recovers once the leader acquires a lease.}
    \label{fig:skewness}
    \Description{A chart showing that as workload becomes more skewed, i.e. the Zipf parameter increases from 0 to 2, successful reads per second falls from 3000 to less then 500.}
\end{figure}

We use Q2 parameters with Zipfian skewness $a$ ranging from 0 to 2 across 1000 keys. At $a=0$, keys have equal probability of being read or written; at $a=2$, the hottest key accounts for 61\% of operations. As in Q2, the leader crashes and a new one is elected. It can use an inherited lease to read any data written by the prior leader, except for data affected by log entries in the limbo region (Section \ref{sec:inherited-lease-reads}). Higher skew makes reads more likely to conflict with limbo region entries. We place 100 log entries into the limbo region and measure the simulated read throughput on the new leader (Figure \ref{fig:skewness}). The number 100 is chosen to stress-test the protocol for skewness effects; in reality the number of limbo entries is determined by workload intensity and the network latency among servers. Once the inherited lease expires, all reads are permitted.

\section{LogCabin evaluation}
\label{sec:eval}

\begin{figure}
    \includegraphics[width=1\linewidth]{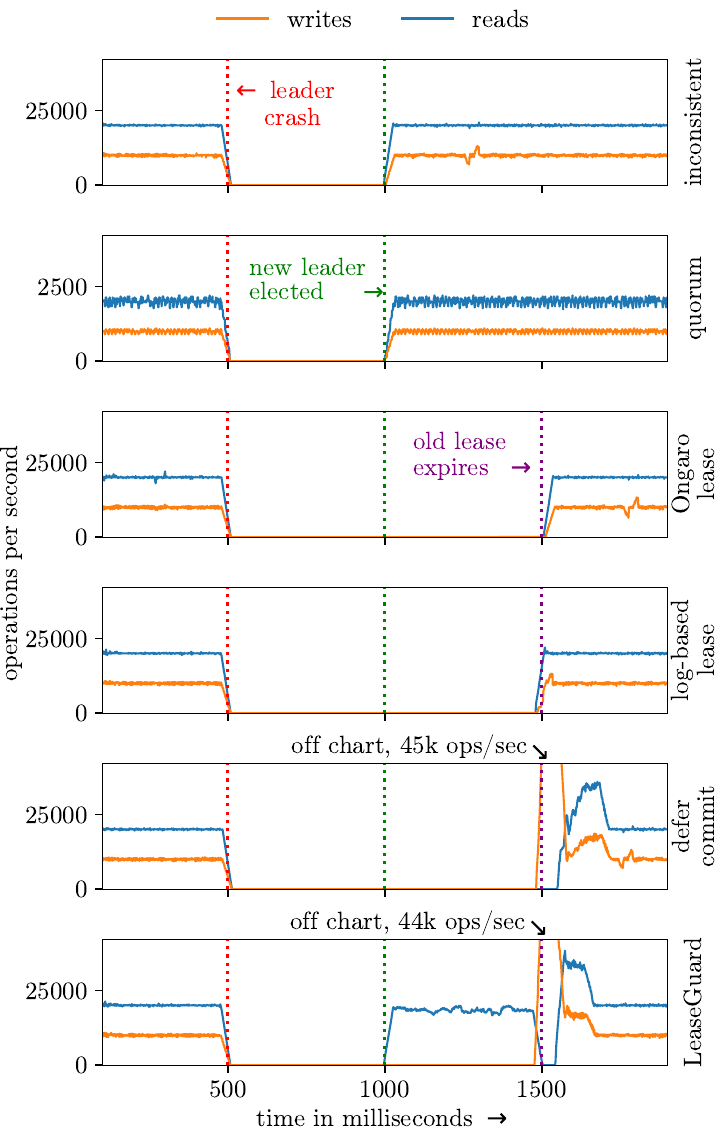}
    \caption{Availability (LogCabin). Ongaro leases and our log-based leases (without optimizations) have the same availability. Our deferred commit optimization improves write throughput when the new leader acquires a lease, and our inherited lease reads restore read availability sooner.}
    \label{fig:availability-logcabin}
    \Description{Six stacked line charts, with a shared horizontal time axis measured in milliseconds since the start of the experiment. The six configurations and their availability characteristics are described in the text.}
\end{figure}

\subsection{Experiment design}
\label{sec:experiment-design-logcabin}

We implemented LeaseGuard in the LogCabin codebase, the reference implementation of Raft. Prior to our work, LogCabin did not use leases: instead, LogCabin leaders ensured consistent reads with a quorum check, as described in the Raft paper. LogCabin implements a key-value database atop the Raft log. Its codebase is 27,000 lines of C++ excluding tests and comments; LeaseGuard required adding or changing 1600 lines. To execute a sufficiently intense open-loop workload from a single client node required enhancing the LogCabin client with an async API, another 650 added or changed lines. The client changes were more difficult than the server changes, but necessary to ensure a valid benchmark, where the client's offered load always matched our intended intensity, no matter whether the servers experienced high latency or hit a throughput ceiling~\cite{closedvsopen}.

We ignore efficiency in our abstract description of the LeaseGuard algorithm, but our C++ implementation is reasonably optimized.
For example, in our pseudocode for CommitEntry (Figure~\ref{fig:leaseguard-logic}), the leader determines if it has any previous-term log entry < $\Delta$ old by iterating its whole log. In C++, the leader caches a \texttt{lastEntryInPreviousTermIndex} variable, which always points  to the latest log entry in the last term, so the CommitEntry check is constant-time. Our pseudocode for ClientRead does not specify how the leader checks if any limbo region entry affects the queried key. In C++, we optimize this check, while preserving Raft's layer separation: the consensus layer manages the log and elections, while the state machine layer manages the keys and values, and the two layers are ignorant of each other's domains. In our implementation, when a node is elected, its consensus layer calls a new method \texttt{StateMachine::setLimboRegion(vector<Entry>)}. This method receives a list of entries in the limbo region and stores an \texttt{unordered\_set<string>} of keys affected by limbo region entries. While the node waits to establish a lease, the state machine efficiently rejects queries involving keys in this set. When the node acquires a lease, the consensus layer calls \texttt{setLimboRegion} again with an empty vector, indicating that the state machine can read all keys freely.

We also implemented in LogCabin the ``Ongaro lease'' protocol, for comparison. By ``Ongaro lease'' we mean the mechanism proposed by Ongaro in~\cite{ongaroThesis}~\S6.4.1. Ongaro sketches the mechanism in a few sentences and has not implemented it, so we have inferred some details and implemented the highest-availability version of the protocol we can imagine. This protocol depends on the Raft rule that a follower will not vote for a candidate if it has heard from a leader less than $\Etime$ ago. In our implementation of Ongaro leases, the leader tracks the start time of each AppendEntries message it sends to any follower. If the follower replies, the leader updates a local variable $s_i$, which is the start time of the last successful AppendEntries sent to node $i$. The leader considers its own $s_i$ to be the current time. If a majority of $s_i$ values are less than $\Etime$ old, then the leader has a lease, because it knows a majority of followers have heard from it recently and therefore have not voted for another node in an election. A leader with a lease runs queries without communicating with followers. In Ongaro's description the lease is shortened slightly to compensate for clock drift; we did not implement this safety check. The Ongaro lease protocol lacks the defer commit and inherited lease read optimizations, and does not allow $\Etime$ to be different from $\Delta$.

In our experiments each node is an EC2 m7g.xlarge in the us-east-1 region. The single client and three servers are in the same placement group. Ping times among them average 155\textmu s. Starting in 2024, servers of this instance type use AWS TimeSync by default, but for maximum clock precision we installed the Precision Time Protocol kernel module, recompiled Amazon's Elastic Network Adapter with physical hardware clock support, and installed Amazon's open-source clock-bound daemon~\cite{clock-bound}. The clock-bound API provides time intervals spanning less than 100\textmu s (clock error less than 50\textmu s). We used our modified LogCabin implementation to replicate two of our three simulated experiments, plus a fourth:

\noindent
{\bf Q1} (Latency): How do leases affect the latency of linearizable reads and writes?

\noindent
{\bf Q2} (Availability): How do leases, inherited leases, and deferred-commit writes affect availability after a leader crashes?

\noindent
{\bf Q4} (Scalability): What is the max throughput with various consistency mechanisms?

\subsection{Latency}
\label{sec:latency-logcabin}

\begin{figure}
    \centering
    \includegraphics[width=1\linewidth]{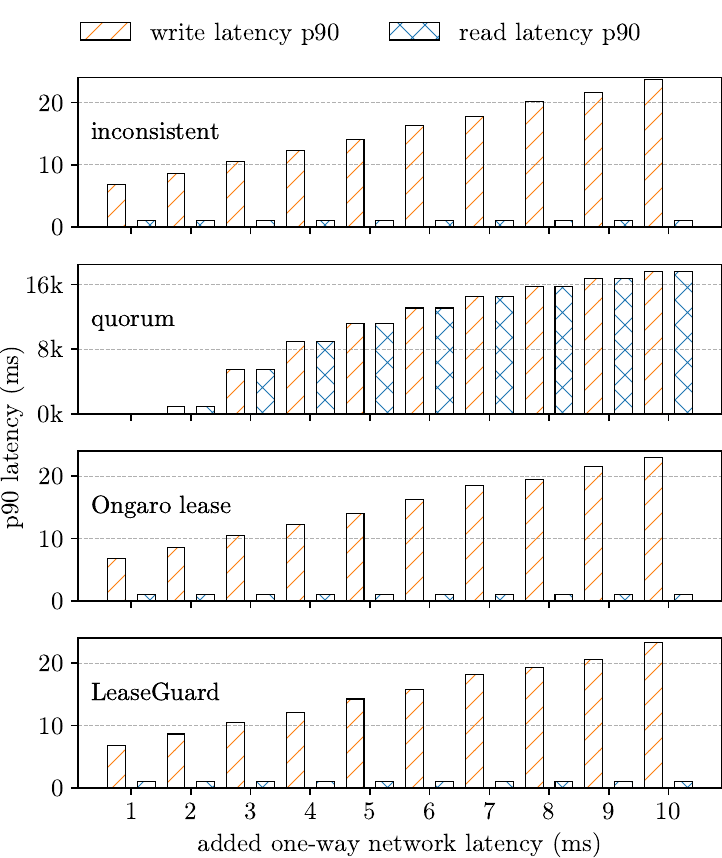}
    \caption{Effect of network latency on read/write latency, milliseconds (LogCabin). Quorum reads hugely increase read and write latency, due to queuing. With Ongaro leases and LeaseGuard, consistent read latency is the same as with no consistency.}
    \label{fig:network-latency-logcabin}
    \Description{Four bar charts showing the effect of network latency on read/write latency, depending on what consistency mechanism we use. If we allow inconsistent reads, then reads are nearly instantaneous in LogCabin, and write latency is a bit more than twice network latency. If we use quorum reads, then both read and write latency are orders of magnitude slower: up to 16 seconds of read and write latency with 10ms of added network latency! With Ongaro leases and LeaseGuard, reads and writes have the same performance as with inconsistent reads.}
\end{figure}

For Q1, the client executes an open-loop workload. One third of operations write a 1-kilobyte value each, and two-thirds of operations read a 1-kilobyte value. The client starts an operation every 300 microseconds and measures its round-trip time. We use ``tc'', the Linux traffic control utility, to add one-way latency from 1 to 10ms between server nodes. This range covers the latencies we would observe between servers within and across regions. Client-server latency is unaffected. Figure~\ref{fig:network-latency-logcabin} shows how network latency affects 90\textsuperscript{th}-percentile read and write latency, in the same three configurations as we used for the simulated experiment, plus ``Ongaro lease''. For all configurations, writes require leader-follower communication and write latency is affected by latency between servers. In the ``inconsistent'', ``Ongaro lease'', and ``LeaseGuard'' configurations, reads require no communication with the followers, so 90\textsuperscript{th}-percentile latency is in the hundreds of microseconds. Performance is similar to the simulation's. In the ``quorum'' configuration, reads require leader-follower communication, and they compete with writes for I/O, harming latency for both. A small increase in network latency causes a huge increase in operation latency due to queueing effects, which we do not observe in our simulation. LeaseGuard permits consistent reads and writes with zero overhead compared to Ongaro leases or the inconsistent configuration.

\subsection{Availability}
\label{sec:availability-logcabin}

As we did in the simulated experiment, we set lease duration $\Delta$ to 1 second (twice $\Etime$) to observe the effect of various consistency mechanisms on availability after an election (Figure~\ref{fig:availability-logcabin}). Ongaro leases do not have a separate $\Delta$ parameter so we set $\Etime$ to 1 second for Ongaro leases. An open-loop workload generator starts 30 operations per millisecond. One-third of operations write 1 kilobyte to a random key, two-thirds read a random key. Keys are randomly selected from a list of 1000 according to a Zipf distribution with $a=0.5$, so the hottest key is chosen 1.6\% of the time. We test four consistency configurations:

\textbf{Inconsistent:} No mechanism ensures linearizability. The same as in the simulation, throughput recovers soon after the election.

\textbf{Quorum:} LogCabin's default consistency mechanism. Note that the Y axis for this chart is one tenth as high as the others: quorum checks dramatically harm throughput. LogCabin shows a much worse impact from quorum checks than our simulation does.

\textbf{Ongaro Lease:} The leader serves reads if a majority of successful RPCs began less than $\Etime$ ago. Throughput is restored when a new leader is elected.

\textbf{Log-based lease:} Our lease protocol with no optimizations. As in the simulation, the system is unavailable after the election until the old leader's lease expires.

\textbf{Defer commit:} LeaseGuard with our write optimization. We see a write spike when the new leader acknowledges all the writes it had buffered while waiting for a lease. This spike is not as vertical as in the simulation, but still goes off the chart to 90 writes per millisecond. The avalanche of acks briefly slows read throughput, which then spikes and recovers. The deferred commit write optimization reduces failed writes and allows the system to stabilize quickly after an election.

\textbf{LeaseGuard:} Our lease protocol with both optimizations (deferred commit writes and inherited lease reads). Unlike our simulation, our actual test produces a significant limbo region of 37 possibly-committed log entries on the new leader that prevent some linearizable reads. Therefore read throughput is slightly lower while the new leader waits for a lease: the client attempts 10,000 reads during this half-second period, but only 9,930 succeed. Nevertheless, the inherited lease read optimization improves read availability while the new leader waits for a lease.

\subsection{Scalability}
\label{sec:latency-logcabin}

\begin{figure}
    \centering
    \includegraphics[width=1\linewidth]{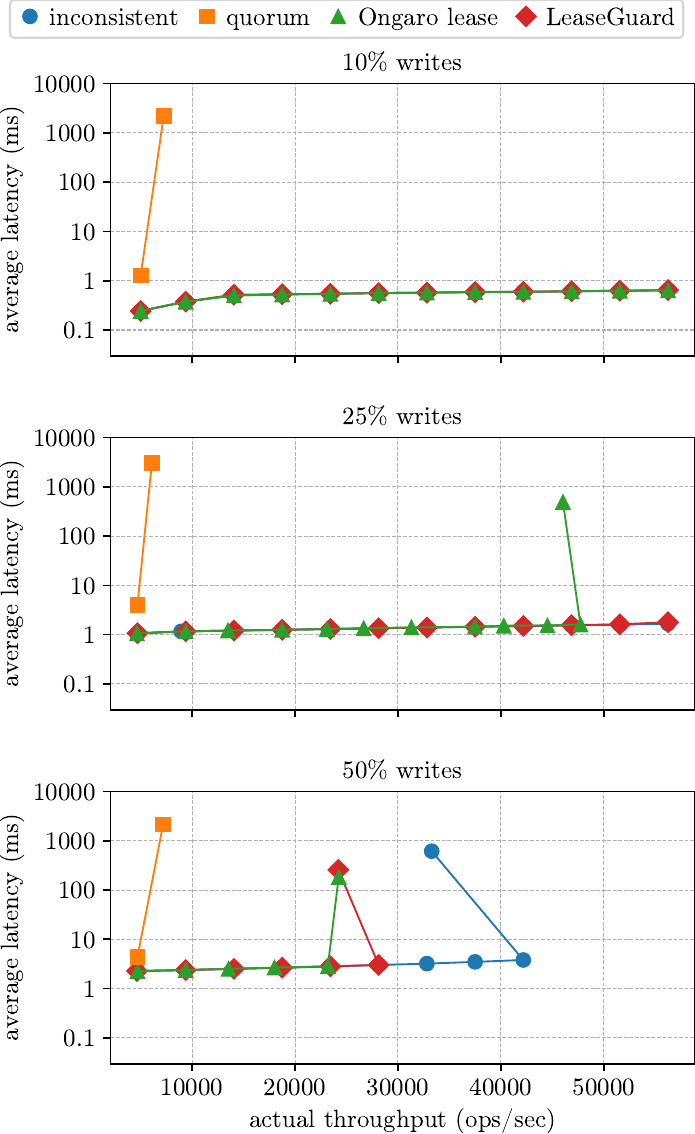}
    \caption{LogCabin scalability with various consistency mechanisms. LeaseGuard is more scalable than quorum check or Ongaro leases and nearly equals the inconsistent configuration.}
    \label{fig:latency-vs-throughput-logcabin}
    \Description{Three line graphs showing throughput vs. latency for a read-only workload and a 25\% and 50\% write workload. For read-only, latency rises after 30,000 operations per second for quorum check and Ongaro lease configurations, but latency stays flat for inconsistent and LeaseGuard. For 25\% reads, latency rises after only 5000 operations per second for quorum check, after 45,000 operations per second for Ongaro lease and LeaseGuard, and stays flat for the inconsistent configuration. For 50\% writes, latency rises after only 5000 operations per second for quorum check, after 25,000 operations per second for Ongaro lease and LeaseGuard, and after 40,000 operations per second for the inconsistent configuration.}
\end{figure}

We subjected LogCabin to increasing load with our async client, starting at 5000 operations per second and stopping at 60,000 operations per second, or once latency increased over 100 ms. As in our latency experiment, the client reads and writes 1 kilobyte values to uniformly-distributed random keys. Figure~\ref{fig:latency-vs-throughput-logcabin} shows LogCabin's throughput and latency with various consistency mechanisms and write ratios. Its throughput with LeaseGuard matches or exceeds that of Ongaro leases for each workload.
\section{Related work}
\label{sec:relwork}

\subsection{Leases}

Leases were originally introduced for distributed file cache consistency~\cite{leases89,liskovClocks}. Consensus with leader leases was implemented for MultiPaxos in ~\cite{paxosLive,chubby,spanner}. Unlike Raft, MultiPaxos does not guarantee Leader Completeness, so it is incompatible with LeaseGuard's simplification (``the log is the lease'') and its read optimization.

A Raft lease protocol was described in~\cite{ongaroThesis}, but not specified in detail or built into LogCabin, Raft's reference implementation. In that protocol, the leader starts a timer, sends a heartbeat message to all nodes, and acquires a lease once a majority replies. The lease lasts until the timer, plus some clock-drift compensation, exceeds some timeout. The leader extends its lease periodically. Followers promise not to vote for a new leader until the last known lease expires. Compared to LeaseGuard, this protocol delays elections and thus harms read and write availability. It is prone to gray failures, where a leader can maintain its lease although it cannot write log entries. It also risks a thundering herd of requests overwhelming the new leader once it acquires a lease.

TiKV uses a separate lease management service called the TiDB Placement Driver, requiring additional infrastructure and communication~\cite{tikv}. The TiDB Placement Driver includes a Raft implementation with a 10-second leader lease, resulting in a 10-second outage after any leader failure~\cite{timestampAsAService}. Spanner implements Paxos with leases using tightly synchronized clocks~\cite{spanner}. In both protocols, a node that knows of a lease does not call an election or cast a vote until that lease expires. In contrast, LeaseGuard does not delay elections or modify the election protocol at all.

YugabyteDB's implementation of Raft leases~\cite{yugabyteLeases} allows leader election to overlap with an outstanding lease, like ours. But during elections, voters must tell the candidate explicitly about existing leases. LeaseGuard is simpler. It avoids the need for an explicit lease-learning mechanism by inferring the lease from the log; this technique was partly inspired by~\cite{cobbs}.

In CockroachDB's Raft variant~\cite{crdb}, replicas hold leases on data ranges, and leaseholders need not be leaders. A replica acquires/extends its lease on a data range by writing a special lease-acquisition log entry. In LeaseGuard, \textit{every} log entry acquires/extends the lease, obviating the need for a special entry type. CockroachDB lacks our improvements to read and write availability.

\subsection{Other mechanisms for consistent reads}

Ongaro proposed a mechanism for ensuring causal consistency in Raft in~\cite{ongaroThesis}~\S6.4.1: the leader includes its lastApplied index with each reply, and the client remembers this index and includes it with each request. If a leader receives a request with a lastApplied greater than its own, it waits until its lastApplied catches up before executing the request. This guarantees causal consistency because Raft ensures that all servers apply the same commands in the same order (State Machine Safety). We implemented this mechanism in MongoDB years ago, but customers dislike it. It requires an application to use the same client for all related requests, which is onerous in a scaled-out cloud application. Besides, some applications require linearizability; causal consistency is relatively weak.

Prior research investigated linearizable reads from \textit{followers} to ease the leader's load. In the Paxos Quorum Lease protocol~\cite{PQL}, followers have leases, so an individual follower can provide linearizable reads without communicating with its peers. The leader needs acknowledgment from each leaseholding follower for each write, so the loss of any single follower can harm write availability. 
Other work~\cite{paxosstore,crdbRead,pqr} ensures linearizable follower reads by contacting a quorum of followers. While these relieve the leader of the read load, they require communication with a majority, and cannot match the efficiency of leader leases.

For strongly-consistent transactional follower reads, Arora et al.~\cite{crdbRead} detect write conflicts using CockroachDB's \textit{write intents}. A write intent is a per-key marker that records any uncommitted update's timestamp, value, and transaction metadata.
A client sends a quorum read to a majority, and followers return data and the timestamp for each key. A follower that encounters a write intent, however, returns only a timestamp without data, signaling an uncommitted update. The client checks whether the highest timestamp in the quorum lacks data, and if so it retries its query. This protocol improves steady-state read scalability by safely offloading reads to followers. 
It bears a resemblance to LeaseGuard's limbo region: both are ways to determine which keys are safe to read. But our limbo region's purpose is to guarantee non-transactional consistent local reads on the \textit{leader} just after an election, not the follower; and the limbo region is a range of log entries, not a marker on a key.

BUC-Raft~\cite{wang2023rethink} proposes two Raft optimizations to reduce latency. In the original Raft protocol, a command must be committed, then applied to the leader's data, before it is acknowledged to the client. BUC-Raft introduces Commit Return (CR), which lets the leader acknowledge writes as soon as they are committed, without awaiting application. This introduces a risk of reading stale data, which BUC-Raft solves with Read Acceleration (RA). RA executes each query on the leader's current data, then patches up the query result with any relevant committed-but-unapplied commands, to ensure freshness.
BUC-Raft claims to preserve linearizability, but this implicitly requires leader leases, since BUC-Raft has no mechanism of its own to prevent clients querying a deposed leader.
RA resembles our limbo-read mechanism, but with fundamental differences. RA aims to reduce steady-state read latency caused by the lag between committing and applying writes, whereas our limbo reads improve availability just after an election.
BUC-Raft's CR and our deferred-commit optimization solve different problems with different techniques: CR acknowledges committed commands sooner and shifts risk to the read path, whereas LeaseGuard accepts writes sooner but withholds acknowledgment until they are committed. 
Overall, BUC-Raft targets steady-leader performance, whereas LeaseGuard ensures consistency and maximizes availability during leader transitions. The approaches are orthogonal and can be composed.

Recent work by Katsarakis et al.~\cite{katsarakis2025law} proves the LAW impossibility: {\em in linearizable asynchronous read/write registers tolerating a single crash, no reads can be local.} They then introduce almost-local reads (ALRs): eager and lazy schemes that execute reads locally but add a lightweight ``sync'' per batch (or piggyback on a timely write) so the result is linearizable under asynchrony.
LeaseGuard, on the other hand, assumes partial synchrony and the availability of bounded-uncertainty clocks. LeaseGuard provides linearizable reads from the leader with no communication overhead or waiting.

\subsection{Bounded-uncertainty clocks}

The use of bounded-uncertainty clocks in distributed systems has become more common recently. Spanner~\cite{spanner} famously uses Google's proprietary time infrastructure to enforce consistency. CockroachDB~\cite{crdb} uses a similar mechanism adapted for public clouds. Accord~\cite{accord} and Detock~\cite{detock} use time to order conflicting transactions. YugabyteDB~\cite{yugabyteTimeSync}, Aurora Limitless~\cite{auroraLimitless}, and Aurora DSQL~\cite{dsql} use AWS TimeSync to enforce consistency with high-precision clocks.
\section{Conclusion}
\label{sec:concl}

We presented a novel leader lease protocol for the Raft consensus algorithm that relies on bounded-uncertainty clocks and the guarantees provided by the Raft log. LeaseGuard simplifies the design and implementation of leader leases while minimizing read and write unavailability during elections. We provide TLA+ specifications and correctness proofs for our protocol. Through analysis, simulation, and benchmarking, we demonstrated LeaseGuard's availability and efficiency benefits.

\clearpage
\bibliographystyle{ACM-Reference-Format}

\end{document}